\documentclass[12pt]{article}

\usepackage{amssymb}
\usepackage{amsmath}
\usepackage{amscd}
\usepackage{latexsym}
\usepackage{cite}

\newcommand{\be}{\begin{equation}}
\newcommand{\ee}{\end{equation}}

\newcommand{\dlt}{\delta}
\newcommand{\prt}{\partial}
\newcommand{\br}{{\bf r}}
\newcommand{\bk}{{\bf k}}
\newcommand{\bfe}{{\bf e}}
\newcommand{\ba}{{\bf a}}
\newcommand{\bp}{{\bf p}}
\newcommand{\bu}{{\bf u}}
\newcommand{\bt}{\beta}
\newcommand{\vp}{\varphi}

\newcommand{\al}{\alpha}
\newcommand{\ra}{\rightarrow}

\newcommand{\gm}{\gamma}
\newcommand{\om}{\omega}
\newcommand{\Om}{\Omega}

\newcommand{\dgr}{\dagger}

\newcommand{\cD}{{\cal D}}
\newcommand{\cX}{{\cal X}}
\newcommand{\cH}{{\cal H}}
\newcommand{\cF}{{\cal F}}

\newcommand{\rgl}{\rangle}
\newcommand{\lgl}{\langle}
\newcommand{\wdH}{\widetilde H}

\begin{document}

\begin{center}

{\Large{\bf Systems with Symmetry Breaking and Restoration} \\ [5mm]

V.I. Yukalov} \\ [3mm]

{\it Bogolubov Laboratory of Theoretical Physics, \\
Joint Institute for Nuclear Research, \\ 
Dubna 141980, Russia; \\
E-mail: yukalov@theor.jinr.ru}

\end{center}

\vskip 3cm

\begin{abstract}

Statistical systems, in which spontaneous symmetry breaking can be
accompanied by spontaneous local symmetry restoration, are considered.
A general approach to describing such systems is formulated, based on
the notion of weighted Hilbert spaces and configuration averaging. The
approach is illustrated by the example of a ferroelectric with mesoscopic
fluctuations of paraelectric phase. The influence of the local symmetry
restoration on the system characteristics, such as sound velocity and
Debye-Waller factor, is discussed. 
\end{abstract}

\vskip 3cm

{\parindent=0pt
{\bf Keywords}: spontaneous symmetry breaking; spontaneous symmetry restoration;

weighted Hilbert spaces; ferroelectric-paraelectric phase transition,
relaxor ferroelectrics 

\vskip 1cm


}

\newpage

\section{Introduction}

It is generally accepted that phase transitions are related to symmetry
changes, so that the transformation from a disordered to an ordered phase
is accompanied by spontaneous symmetry breaking \cite{1,2,3}. There also
exist more complicated cases, when the symmetry, being broken in the
major part of the sample, at the same time, is restored in other parts
of the same system, which is called symmetry restoration \cite{4}.
Reciprocally, in the generally disordered phase, there can appear local
regions with broken symmetry. In dynamical modeling, the arising regions
of symmetry that differs from the symmetry of the surrounding matrix are
related to the generation of solitonic clusters and local coherent
structures \cite{5,6,7,8,9,10,11,12,13,14,15,16,17,18,19}. It turns out that the systems with such
local structures are more stable, as compared to homogeneous systems.
Generally, the systems that become more stable by spontaneously changing
their properties pertain to the class of self-optimizing systems \cite{20}.
There are numerous examples of condensed matter, where the sample is not
homogeneous, but consists of regions of different symmetry. One tells
that such systems display mesoscopic phase separation, and they are
termed heterophase. This, for instance, concerns many high-temperature
superconductors, in which superconducting regions coexist with normal
regions \cite{21,22,23,24,25}, and some low-temperature superconductors
\cite{26,27}. Such a coexistence of superconducting and normal phases
can occur even in atomic nuclei \cite{28}. In many magnetic materials,
magnetically ordered phase includes paramagnetic clusters
\cite{29,30,31,32,33}. Around the points of structural phase transitions,
there exist regions, where phases with different symmetry coexist
\cite{34,35,36,37,38,39,40}. A number of ferroelectric materials displays the
coexistence of ferroelectric and paraelectric phases \cite{41,42,43,44,45,46,47,48,49,50,51,52,53,54},
which also concerns such novel materials as relaxor ferroelectrics
\cite{55,56,57,58}. Much more examples of heterophase matter can be
found in the review articles \cite{59,60}.

The basic difficulty in the description of heterophase systems is the
necessity of dealing with two or more different phases, possessing
principally different symmetry properties, but coexisting inside the
same volume. The problem is aggravated by the fact that the location
of the germs of different phases in space is chaotic. The situation
is drastically different from the case of a sample consisting of several
domains with well defined spatial locations and structure \cite{61}. A
typical heterophase system is a mixture of regions, randomly located in
space and having various and often very irregular shapes. Moreover, in
many cases, the heterophase regions are not static, but can move in space,
vary their shapes, and even appear and disappear.

To describe such a complicated matter, it has been necessary to develop an
approach allowing for the treatment of these inhomogeneous and
nonequilibrium systems. More precisely, such systems are to be locally
equilibrium, since the notion of phase requires the existence of at least
local equilibrium. A general approach for treating heterophase systems has
been advanced \cite{62,63,64,65,66} and summarized in reviews \cite{59,60}.

One of the main problems in treating the systems with coexisting regions
of different symmetry is how to separate the states, corresponding to
different symmetry properties, in the space of microscopic quantum states.
The standard situation is when the system as a whole is characterized by
a given Hilbert space, with a prescribed symmetry. Then how would it be
possible to describe subsystems of different symmetry in the frame of the
same Hilbert space? This can be done, for instance, by imposing different
conditions on the equations characterizing different phases \cite{62,67}
or by invoking the method of restricted trace \cite{68}. Probably, the most
powerful, convenient, and rigorous method of separating different phases
is the approach based on the notion of weighted Hilbert spaces, advanced
in Refs. \cite{59,60}. In the latter publications, however, this method
was not formulated in the most mathematically general way. The aim of the
present paper is to further develop the method of weighted Hilbert spaces,
formulating it in the most general form. As an illustration, the method
will be applied to deriving a general model of a heterophase ferroelectric.
The choice of ferroelectrics for illustration is caused by several reasons.
First, there are numerous examples of the heterophase systems of this type
\cite{41,42,43,44,45,46,47,48,49,50,51,52,53,54,55,56,57,58,59,60}, hence their correct description is of great importance.
Second, the general model related to these materials is generic for many
other substances, hence it serves as a good example for extending the
approach to other heterophase systems.

\section{Single-Phase Systems}

Before developing an approach for treating multiphase systems, it is
necessary to briefly recall the main points of describing the standard case
of a single-phase system, which will be used in what follows. For
generality, quantum systems are considered throughout the paper.

Let us have a set $\{\varphi_n\}$ of states forming a basis for the closed
linear envelope
\be
\label{1}
\overline{\cal E} \equiv {\rm Span}_n \{ \vp_n \} \; .
\ee
Here the index $n$ implies a multi-index that can pertain to either
discrete or continuous set. The discrete set can be infinite.
Let the scalar product $<f|h>$ be defined for each pair $f,h\in\cal{E}$.
The norm, generated by the scalar product, is
\be
\label{2}
|| f || \equiv \sqrt{ \lgl f \; | \; f \rgl } \qquad
( f \in  \overline{\cal E} ) \;  .
\ee
Completing the linear envelope (\ref{1}) by the norm yields the complete
normed space, that is, the Hilbert space
\be
\label{3}
{\cal H} \equiv \{\; \overline{\cal E} , \; || f || \; \} \;  .
\ee
The so-defined Hilbert space can be separable or not, depending on the
physics of a concrete problem and, respectively, on the nature of the
multi-index $n$ enumerating the basis $\{\varphi_n\}$. The existence of
a Hilbert space, associated with the considered physical system, is the
necessary prerequisite for characterizing the system.

The basis $\{\varphi_n\}$ can be taken to be orthonormalized, such that
\be
\label{4}
\lgl \; \vp_m \; | \; \vp_n \; \rgl = \dlt_{mn} \; .
\ee
And let it be complete. Then for each $f,h\in \overline{\cal{E}}$, one can write their
expansions over the basis,
\be
\label{5}
f = \sum_n \; f_n \vp_n \; , \qquad h = \sum_n \; h_n \vp_n \; ,
\ee
with the expansion coefficients
$$
f_n = \lgl \; \vp_n \; | \; f \; \rgl \; , \qquad
 h_n = \lgl \; \vp_n \; | \; h \; \rgl \; .
$$
Therefore, the scalar product can be represented as
\be
\label{6}
\lgl \; f\; | \; h \; \rgl = \sum_n \; f_n^* h_n \; .
\ee

Suppose the algebra $\cal{A}$ of local observables, represented by
Hermitian operators $\hat A\in\cal{A}$, be given on $\cal{H}$. The matrix
elements of $\hat A$ over the basis $\{\varphi_n\}$ are
\be
\label{7}
A_{mn} \equiv \lgl \; \vp_m \; | \; \hat A \; | \; \vp_n \; \rgl \;  .
\ee
The system statistics are characterized by a statistical operator $\hat {\rho}$
that is a trace-one operator, acting on $\cal{H}$. The operator averages
are defined as
\be
\label{8}
\lgl \hat A \rgl \equiv {\rm Tr}_{\cal H} \; \hat\rho \hat A =
\sum_n \; \lgl \; \vp_n \; | \; \hat \rho \hat A \; | \; \vp_n \; \rgl \;  .
\ee
The set of all available averages is termed the {\it statistical state}:
\be
\label{9}
\lgl {\cal A} \rgl = \{ \lgl \hat A \rgl \} \;  .
\ee

Characterizing thermodynamics phases, a special role is played by the order
operator $\hat{\eta}\in\cal{A}$, whose average defines the system
{\it order parameter}
\be
\label{10}
\eta = \lgl \;\hat \eta \; \rgl \; .
\ee
In general, this can be a scalar, vector, or matrix quantity, so that one
talks of an order parameter just for short. The main point is that this
order parameter is specific for each thermodynamic phase, such that
different phases, possessing different symmetries, enjoy different order
parameters.

\section{Symmetry Breaking}

In the language of symmetries, the existence of phase transitions assumes
the following. If there is a statistical state that is invariant with
respect to a symmetry group, then it can be decomposed into a sum of
several terms describing different pure phases \cite{1,69,70}. To select a
state with a particular type of symmetry, one employs one of the variants
of the quasiaveraging techniques \cite{1,67,71}. When the state with a broken
symmetry is thermodynamically more stable than the invariant state, one
calls this the spontaneous breaking of symmetry.

The methods of quasiaverages allow one to select a particular state with
the desired symmetry only in the case of thermodynamically equilibrium
systems, when all the system is characterized by one and the same symmetry.
But our aim is to describe the situation, when inside the considered
system there appear regions with different types of symmetry. How could
we proceed in such a principally different case? For this purpose, it is
necessary to develop a more general method of symmetry breaking, which
could be used not only for equilibrium systems, but also for
quasiequilibrium, metastable, or even for arbitrary nonequilibrium systems.
Such a general method is developed below.

Let us consider a statistical system that, under different thermodynamic
conditions, could be in different thermodynamic phases, enumerated by the
index $\nu = 1,2,\ldots$. We keep in mind that the space of microscopic
states, related to the considered system, is the Hilbert space (\ref{3}),
with a basis $\{\varphi_n\}$, as described in Sec. 2. Let us put into
correspondence to a vector $\varphi_n$ a probability
\be
\label{11}
p_n^\nu \equiv p^\nu(\vp_n) \; .
\ee
The set $\{p_n^\nu\}$ of these probabilities forms a probability measure
with the standard normalization property
\be
\label{12}
\sum_n \; p_n^\nu = 1 \qquad (0 \; \leq \; p_n^\nu \; \leq 1 ) \;  .
\ee

Now, let us introduce the {\it weighting operator}
\be
\label{13}
\hat W_\nu \equiv
\sum_n \; p_n^\nu \; | \; \vp_n \rgl \lgl \vp_n \; |  \; .
\ee
Using this, we define the {\it weighted scalar product}
\be
\label{14}
\lgl \; f \; | \; h \; \rgl_\nu \equiv
\lgl \; f\; | \; \hat W_\nu \; | \; h \; \rgl \; .
\ee
With the weighting operator (\ref{13}), we have
\be
\label{15}
\lgl \; f \; | \; h \; \rgl_\nu =
\sum_n \; p_n^\nu \; \lgl \; f \; | \; \vp_n \; \rgl
\lgl \; \vp_n \; | \; h \; \rgl  \;  .
\ee
Under expansions (\ref{5}), the latter reads as
\be
\label{16}
\lgl \; f \; | \; h \; \rgl_\nu = \sum_n \; p_n^\nu f_n^* h_n \;  .
\ee
The scalar product (\ref{15}) generates the norm
\be
\label{17}
|| f ||_\nu \equiv
\sqrt{ \lgl \; f \; | \; f \; \rgl_\nu } \qquad
( f \in  \overline{\cal E} )   \;  ,
\ee
which, in view of form (\ref{16}), can be written as
\be
\label{18}
|| f ||_\nu \equiv
\sqrt{ \sum_n \; p_n^\nu \; | f_n|^2 } \; .
\ee

The closed linear envelope (\ref{1}), equipped with norm (\ref{17}), is the
{\it weighted Hilbert space}
\be
\label{19}
{\cal H}_\nu \equiv \{ \overline{\cal E} ,\; || f||_\nu \}  \;  .
\ee

The representation $\cal{A_\nu}$ of the algebra of local observables
$\cal{A}$, acting on the weighted Hilbert space (\ref{19}), consists
of the operators $\hat{A}_\nu$ defined through their matrix elements
\be
\label{20}
\lgl \; \vp_m \; | \; \hat A_\nu \; | \; \vp_n \; \rgl \equiv
\lgl \; \vp_m \; \left | \; \frac{1}{2} \left [ \hat A, \; \hat W_\nu
\right ]_+ \; \right | \; \vp_n \; \rgl \; .
\ee
Taking into account the weighting operator (\ref{13}) gives
\be
\label{21}
\lgl \; \vp_m \; | \; \hat A_\nu \; | \; \vp_n \; \rgl =
\frac{1}{2} \left ( \; p_m^\nu + p_n^\nu \; \right )
\lgl \; \vp_m \; | \; \hat A \; | \; \vp_n \; \rgl  \;  .
\ee

The operator averages are
\be
\label{22}
\lgl \hat A_\nu \rgl \equiv {\rm Tr}_{{\cal H}_\nu} \left ( \;
\hat\rho \hat A \; \right )_\nu =  \sum_n \; \lgl \; \vp_n \; | \;
\left ( \hat\rho \hat A \right )_\nu \; | \; \vp_n \; \rgl \;  .
\ee
With the matrix elements (\ref{21}), we get
\be
\label{23}
\lgl \hat A_\nu \rgl = \sum_n \; p_n^\nu \;
\lgl \; \vp_n \; | \; \hat\rho \hat A \; | \; \vp_n \; \rgl  \;  .
\ee

Similarly, the average of the order operator $\hat{\eta}_\nu\in\cal{A}_\nu$
becomes
\be
\label{24}
\lgl \; \hat\eta_\nu \; \rgl = \sum_n \; p_n^\nu \;
\lgl \; \vp_n \; | \; \hat\eta \; | \; \vp_n \; \rgl  \;  .
\ee
Respectively, the order parameter is
\be
\label{25}
\eta_\nu \equiv \lgl \; \hat\eta_\nu \; \rgl \; .
\ee
The probability measure $\{p_n^\nu\}$ is to be defined so that to guarantee
the order parameter, specifying the chosen thermodynamics phase. This means
that the probabilities (\ref{11}) should select the states typical of the
considered phase \cite{72,73}, which does not need to be equilibrium. By this
construction, it is clear that the scheme, based on the weighted Hilbert
spaces, includes as a particular case the selection of phases by means of
the quasiaveraging method, since the latter also chooses the states typical
of the desired phase, but provided this phase corresponds to a stable
equilibrium system.

\section{Multiphase Systems}

First of all, let us stress the difference of the case we try to describe,
as compared to the Gibbs phase mixture \cite{74}. In the latter case,
the system is spatially separated into several {\it macroscopic} regions
filled by different thermodynamics phases. But in the case we are
interested in, the system is a {\it heterophase} mixture, comprising
the {\it mesoscopic} germs of several thermodynamic phases, which are
randomly intermixed and coexist in a region of thermodynamic parameters
in the same volume \cite{59,60}. The space of states for such a heterophase
system is the {\it fiber space}
\be
\label{26}
\widetilde{\cal F} \equiv \bigotimes_\nu \; {\cal H}_\nu\;  .
\ee
The basis of the latter, $\{\widetilde \varphi_{\bf n}\}$, is made of the
tensor products
\be
\label{27}
\widetilde\vp_{\bf n} \equiv \bigotimes_\nu \; \vp_{n_\nu} \;  ,
\ee
in which the notation for the multi-index
$$
{\bf n} \equiv \{ n_1,\; n_2, \; n_3, \ldots \}
$$
is used. Any state $\widetilde f\in\widetilde{\cal F}$ can be decomposed over this
basis:
$$
\widetilde f = \sum_{ {\bf n} } \; f_{\bf n} \widetilde\vp_{\bf n}
\qquad \left ( \; f_{\bf n} = \lgl \; \widetilde\vp_{\bf n} \; | \; \widetilde f \;
\rgl \; \right ) \; .
$$
Then the scalar product of any pair
$\widetilde f, \widetilde h\in\widetilde{\cal{F}}$ is given by
\be
\label{28}
\lgl \; \widetilde f \; | \; \widetilde h \; \rgl = \sum_{\bf n} \;
f_{\bf n}^* h_{\bf n}  \;  .
\ee
The scalar product generates the norm
\be
\label{29}
|| \widetilde f || \equiv
\sqrt{\; \lgl \; \widetilde f \; | \; \widetilde f\; \rgl } \; .
\ee
The fiber space (\ref{26}) can be equivalently represented as the closed
linear envelope
\be
\label{30}
\widetilde{\cal F} = \left \{ {\rm Span}_{\bf n} \{ \widetilde\vp_{\bf n} \} , \;
|| \widetilde f || \right \} \; ,
\ee
equipped with norm (\ref{29}).

The operators of observables $\widetilde{A}$, acting on $\widetilde{\cal F}$, are
defined as the direct sums
\be
\label{31}
\widetilde A = \bigoplus_\nu \; \hat A_\nu \;  .
\ee
The related matrix elements are given by the expressions
\be
\label{32}
\lgl \; \widetilde\vp_{\bf m} \; | \; \widetilde{A} \; | \;
\widetilde\vp_{\bf n} \; \rgl = \sum_\nu \; \lgl \; \vp_{m_\nu} \; | \;
\hat A_\nu \; | \; \vp_{n_\nu} \;
\rgl  \prod_{\mu(\neq\nu)} \dlt_{m_\mu n_\mu}  \; .
\ee

The operator averages are defined as
\be
\label{33}
\lgl \widetilde A \rgl \equiv {\rm Tr}_{\widetilde{\cal F}} \;
\widetilde\rho \widetilde A = \sum_{\bf n} \;
\lgl \widetilde\vp_{\bf n} \; | \; \widetilde\rho \widetilde A \; | \;
\widetilde\vp_{\bf n} \; \rgl \; .
\ee
This, in view of form (\ref{27}), yields
\be
\label{34}
\lgl \widetilde A \rgl = \sum_\nu \sum_n \; \lgl \; \vp_n \; | \;
( \hat\rho \hat A )_\nu  | \; \vp_n \; \rgl  \;  .
\ee
Comparing this with (\ref{22}), we obtain
\be
\label{35}
\lgl \widetilde A \rgl = \sum_\nu \; \lgl \widetilde A_\nu \rgl  \; .
\ee
The set $\{<\widetilde{A}>\}$ of all observable averages is the statistical
state of the heterophase system.

\section{Phase Configurations}

When the system is inhomogeneous, being composed of many mesoscopic regions
of different phases, we need, first of all, to describe the spatial
distribution of these regions inside the system. For this purpose, the space
$\mathbb{V}$, occupied by the system, can be decomposed into the subregions,
whose set $\{\mathbb{V}_\nu\}$ forms an orthogonal covering:
\be
\label{36}
\mathbb{V} = \bigcup_\nu \mathbb{V}_\nu \; , \qquad
V = \sum_\nu \; V_\nu \; ,
\ee
such that
\be
\label{37}
\mathbb{V}_\mu \; \bigcap \; \mathbb{V}_\nu = \dlt_{\mu \nu}
\mathbb{V}_\nu \; ,
\ee
where
\be
\label{38}
V \equiv \int_\mathbb{V} d\br \; , \qquad
V_\nu \equiv \int_{\mathbb{V}_\nu} d\br \; .
\ee
The set of the regions, occupied by a $\nu$-phase is described by the
manifold characteristic function \cite{75}, or {\it manifold indicator}
\begin{eqnarray}
\label{39}
\xi_\nu(\br) \equiv \left \{ \begin{array}{ll}
1 , & \br \in \mathbb{V}_\nu \\
0 , & \br \not\in \mathbb{V}_\nu \end{array}
\right. \; .
\end{eqnarray}
These functions satisfy the properties
\be
\label{40}
\sum_\nu \; \xi_\nu(\br) = 1 \qquad ( \br \in \mathbb{V} ) \; 
\ee
and
\be
\label{41}
\int_\mathbb{V} \xi_\nu(\br) \; d\br = V_\nu \; .
\ee
The collection of all manifold indicators defines the {\it phase configuration}
\be
\label{42}
\xi \equiv \{ \xi_\nu(\br) : \; \nu = 1, \; 2, \; \ldots ; \;
\br \in \mathbb{V} \} \; .
\ee

In turn, the space, occupied by a $\nu$-phase, can be decomposed into subregions
$\mathbb{V}_\nu$, forming an orthogonal subcovering $\{\mathbb{V}_\nu\}$, such that
\be
\label{43}
\mathbb{V}_\nu = \bigcup_{i=1}^{z_\nu} \mathbb{V}_{\nu i} \; ,
\ee
and
\be
\label{44}
\mathbb{V}_{\mu i} \; \bigcap \; \mathbb{V}_{\nu j} =
\dlt_{\mu \nu} \dlt_{ij} \mathbb{V}_{\nu i} \; .
\ee
Each subregion $\mathbb{V}_{\nu i}$ is described by its manifold indicator
\begin{eqnarray}
\label{45}
\xi_{\nu i}(\br - \ba_{\nu i} ) \equiv \left \{ \begin{array}{ll}
1 , & \br \in \mathbb{V}_{\nu i} \\
0 , & \br \not\in \mathbb{V}_{\nu i} \end{array}
\right. \; ,
\end{eqnarray}
in which $\bf{a}_{\nu i}$ is a fixed vector pertaining to $\mathbb{V}_{\nu i}$.
Then the manifold indicator (\ref{39}) writes as
\be
\label{46}
\xi_\nu(\br) = \sum_{i=1}^{z_\nu} \;
\xi_{\nu i} (\br - \ba_{\nu i} ) \; .
\ee

Under a given configuration, the relative volume, occupied by a $\nu$-phase, is
characterized by its geometrical fraction
\be
\label{47}
x_\nu \equiv \frac{1}{V} \; \int_\mathbb{V} \; \xi_\nu(\br) \; d\br =
\frac{V_\nu}{V} \; .
\ee
As is clear, the latter satisfy the normalization condition
\be
\label{48}
\sum_\nu x_\nu = 1 \qquad ( 0 \; \leq x_\nu \leq 1 ) \; .
\ee
This shows that the set $\{x_\nu\}$ of all admissible geometrical fractions,
with the given normalization condition, can be regarded as a probability measure.

Since the the phase regions are randomly distributed inside the considered
system, the configuration (\ref{42}) is to be treated as a random variable.
The related differential measure can be defined as
\be
\label{49}
\cD \xi = \lim_{\{ z_\nu\ra\infty \} }\; \dlt
\left ( \sum_\nu \; x_\nu - 1 \right )
\prod_\nu d x_\nu \prod_\nu \prod_{i=1}^{z_\nu} \;
\frac{d\ba_{\nu i} }{V} \; .
\ee
All possible configurations constitute a topological space
\be
\label{50}
\cX \equiv \{ \xi, \; \cD\xi \} \; .
\ee

The above constructions describe the situation, when the system is separated
into several regions filled by different phases, and these regions are randomly
distributed in space.

\section{Configuration Averaging}

An inhomogeneous heterophase system is not necessarily in complete equilibrium.
But it must be at least quasi-equilibrium in order that it would be admissible to
talk about the germs of phases. As is known, the notion of phases is not strictly
defined for finite systems. The mathematically rigorous definition of phases
assumes the introduction of thermodynamic limit \cite{2,3,72,74,76}. However, in
practice, it is possible to speak about germs of phases already when each of such
germs consists of a large number of particles $N \gg 1$. Monte Carlo simulations
show that thermodynamics phases can be well defined already for $10-100$ particles
in a finite cluster \cite{77}.  Respectively, though the symmetry, related to a
phase, may be not strictly defined for a finite cluster, but it is possible to talk
about an {\it asymptotic symmetry} that is approximately defined for a large number
of particles $N \gg 1$, keeping in mind that the symmetry becomes exact in the
thermodynamic limit.

A quasi-equilibrium system is described by a {\it quasi-Hamiltonian} \cite{59}
depending on a given phase configuration and having the operator structure as
in (\ref{31}):
\be
\label{51}
\hat Q(\xi) =  \bigoplus_\nu \hat Q_\nu (\xi_\nu) \; .
\ee

The partition function
\be
\label{52}
Z \equiv {\rm Tr}_{\widetilde\cF} \int \exp \left \{ - \hat Q(\xi)
\right  \} \; \cD\xi \;
\ee
includes the quantum averaging over the given quantum variables and the
averaging over phase configurations. This function defines the quasi-equilibrium
thermodynamic potential
\be
\label{53}
y \equiv -\; \frac{1}{N} \; \ln Z \; .
\ee

Considering asymptotically large systems, we, as usual, keep in mind the
thermodynamic limit
\be
\label{54}
N ~ \ra ~ \infty \; , \qquad V ~ \ra ~ \infty \; , \qquad
\frac{N}{V} ~ \ra ~ const \; .
\ee

Similarly to (\ref{31}), the operators of observables, under the given
phase configuration, have the form
\be
\label{55}
\hat A(\xi) = \bigoplus_\nu \hat A_\nu(\xi_\nu) \; .
\ee
And for what follows, it is convenient to introduce the notation
\be
\label{56}
\hat A_\nu(x_\nu) \equiv \lim_{\xi_\nu\ra x_\nu}
\hat A_\nu(\xi_\nu) \; .
\ee

The statistical operator of a multi-phase system, with a fixed phase
configuration, is
\be
\label{57}
\hat\rho(\xi) = \frac{1}{Z} \;
\exp \left \{ - \hat Q(\xi) \right \} \; .
\ee
The observable quantities, related to the operators (\ref{55}), are given
by the averages
\be
\label{58}
\lgl \widetilde A \rgl \equiv {\rm Tr}_{\widetilde\cF}
\int \hat\rho(\xi) \hat A(\xi) \; \cD\xi \; .
\ee
The following theorem is valid \cite{59,78,79}.

\vskip 5mm

{\bf Theorem 1}.

\vskip 3mm

If $\hat {Q}_\nu(\xi_\nu)$ can be represented as an expansion in powers
of $\xi_\nu$, then the thermodynamic potential (\ref{53}), in the
thermodynamic limit (\ref{54}),
asymptotically equals
\be
\label{59}
y ={\rm abs} \; \min_{\bf w} y({\bf w}) \; ,
\ee
with the set
\be
\label{60}
{\bf w} \equiv \{ w_1, \; w_2, \; \ldots \} \;
\ee
forming the probability measure enjoying the standard properties
\be
\label{61}
\sum_\nu \; w_\nu = 1  \qquad ( 0 \; \leq w_\nu \; \leq 1) \; ,
\ee
and where
\be
\label{62}
y({\bf w}) =  \sum_\nu \; y_\nu(w_\nu) \;
\ee
is the sum of the terms
\be
\label{63}
y_\nu(w_\nu) = - \; \frac{1}{N} \; \ln {\rm Tr} \; Z_\nu \; ,
\ee
in which
\be
\label{64}
Z_\nu = {\rm Tr}_{\cH_\nu} \; \exp \left \{ - \hat Q_\nu(w_\nu)
\right \} \; .
\ee

\vskip3mm

The quantities $w_\nu$ are the {\it phase geometric probabilities},
showing the corresponding weights of the coexisting thermodynamic phases.

\vskip 5mm

{\bf Corollary 1}

\vskip 3mm

From this theorem, it follows that the thermodynamic potential (\ref{59})
can also be represented in the form
\be
\label{65}
y = -\; \frac{1}{N} \; \ln {\rm Tr} \; \widetilde Z \; ,
\ee
where
$$
\widetilde Z \equiv \prod_\nu Z_\nu \; .\
$$

\vskip 3mm

The above theorem defines the thermodynamic potential of a heterophase
system. The observable quantities, corresponding to the averages of the
operators from the algebra of local observables, are given by the following
theorem \cite{59,78,79}.

\vskip 5mm

{\bf Theorem 2}.

\vskip 3mm

Assume that $\hat{Q}_\nu(\xi_\nu)$ and $\hat{A}_\nu(\xi_\nu)$ can be expanded in
powers of $\xi_\nu$, then the averages (\ref{58}), for asymptotically large $N$,
take the form of the sum
\be
\label{66}
\lgl \widetilde A \rgl =  \sum_\nu \; \lgl \hat A_\nu \rgl \; ,
\ee
with the terms
\be
\label{67}
\lgl \hat A_\nu \rgl = {\rm Tr}_{\cH_\nu} \; \hat\rho_\nu  \; \hat A_\nu \; ,
\ee
in which
\be
\label{68}
\hat\rho_\nu = \frac{1}{Z_\nu} \;
\exp \left \{ - \hat Q_\nu(w_\nu) \right \} \; ,
\ee
and where the notation $\hat A_\nu \equiv \hat A_\nu(w_\nu)$ is used.

\vskip 5mm

{\bf Corollary 2}.

\vskip 3mm

Equivalently, the averages (\ref{66}) can be represented in another way by
introducing the operators
\be
\label{69}
\widetilde A \equiv \bigoplus_\nu \hat A_\nu \; ,
\ee
for which
\be
\label{70}
\lgl \widetilde A \rgl =
{\rm Tr}_{\widetilde\cF} \widetilde\rho \widetilde A \; ,
\ee
where
$$
\widetilde\rho = \bigotimes_\nu \hat\rho_\nu \; .
$$

\vskip 3mm

The second theorem defines the method of calculating the averages for a
heterophase system.

\section{Effective Hamiltonians}

The quasi-Hamiltonian $\hat{Q}_\nu(\xi_\nu)$ is connected with the
local Hamiltonian $\hat{H}_\nu(\bf r, \xi_\nu)$ through the relation
\be
\label{71}
\hat Q_\nu(\xi_\nu) = \int \bt_\nu(\br,\xi_\nu)
\hat H_\nu(\br,\xi_\nu) \; d\br \; ,
\ee
in which the inverse local temperatures $\beta_\nu(\bf r, \xi_\nu)$ play
the role of the Lagrange multipliers.

If the considered system is such that the phase fluctuations can appear
randomly in any part of the system, the latter is called phase-uniform
on average \cite{59}. For such a system, the configuration averaging of
the inverse temperatures gives
\be
\label{72}
\bt_\nu \equiv \int \bt_\nu(\br,\xi_\nu) \; \cD\xi \; .
\ee
The mixture of different phases is in thermal equilibrium, when the
temperatures of these phases coincide:
\be
\label{73}
\bt_\nu = \bt = \frac{1}{T} \qquad (\forall \nu) \; .
\ee
Then the quasi-Hamiltonian $\hat{Q}_\nu(w_\nu)$, entering the partition
function (\ref{64}), becomes
\be
\label{74}
\hat Q_\nu(w_\nu) = \bt \hat H_\nu \; ,
\ee
with the {\it renormalized Hamiltonian}
\be
\label{75}
\hat{H}_\nu \equiv \int \hat H_\nu(\br,w_\nu) \; d\br \; .
\ee
The partial statistical operators (\ref{68}) acquire the form
\be
\label{76}
\hat\rho_\nu = \frac{1}{Z_\nu} \; \exp \left ( -\bt \hat H_\nu
\right ) \; ,
\ee
with the partition functions
\be
\label{77}
Z_\nu = {\rm Tr}_{\cH_\nu} \; \exp \left ( -\bt \hat H_\nu
\right ) \; .
\ee

It is convenient to introduce the {\it effective Hamiltonian}
\be
\label{78}
\widetilde H \equiv \bigoplus_\nu \hat H_\nu \; ,
\ee
using which, the thermodynamic potential (\ref{65}) can be represented as
\be
\label{79}
y = - \; \frac{1}{N} \; \ln \; {\rm Tr}\;
e^{-\bt\widetilde H} \; .
\ee
The terms of sum (\ref{78}) are called the {\it phase-replica Hamiltonians},
since they have a similar mathematical structure, but are associated with
different phases.

In order to connect the thermodynamic potential (\ref{79}) with the free
energy, let us recall that the latter is defined as
\be
\label{80}
F \equiv - T \ln \; {\rm Tr}\;  e^{-\bt\widetilde H} \; .
\ee
Therefore, the potential $y$ is directly connected to the free energy
by means of the relations
$$
F = NTy = \sum_\nu \; F_\nu \;
$$
and
$$
F_\nu = - T \ln \; {\rm Tr}\; Z_\nu = NTy_\nu \; .
$$

To specify the consideration, let us take the local Hamiltonians in the
usual form
$$
\hat H_\nu(\br,\xi_\nu) = \xi_\nu(\br) \psi^\dgr_\nu(\br)
\left ( - \; \frac{\nabla^2}{2m} + U \right ) \psi_\nu(\br) \; +
$$
\be
\label{81}
+ \; \frac{1}{2} \int \xi_\nu(\br) \xi_\nu(\br') \psi^\dgr_\nu(\br)
\psi^\dgr_\nu(\br') \Phi(\br - \br') \psi_\nu(\br')
\psi_\nu(\br) \; d\br \; .
\ee
Here $U = U(\bf r)$ is an external potential and $\psi_\nu(\bf r)$ are the
field operators of the particles forming the system.

Then the phase-replica Hamiltonians are
$$
\hat H_\nu = w_\nu \int \psi^\dgr_\nu(\br)
\left ( - \; \frac{\nabla^2}{2m} + U \right )
\psi_\nu(\br) \; d\br \; +
$$
\be
\label{82}
+ \; \frac{w_\nu^2}{2} \int \psi^\dgr_\nu(\br) \psi^\dgr_\nu(\br')
\Phi(\br - \br') \psi_\nu(\br')
\psi_\nu(\br) \; d\br d\br' \; .
\ee
Introducing the notation for the kinetic-energy operator
\be
\label{83}
\hat K_\nu \equiv \int \psi^\dgr_\nu(\br)
\left ( - \; \frac{\nabla^2}{2m} + U \right )
\psi_\nu(\br) \; d\br \;
\ee
and for the operator
\be
\label{84}
\hat\Phi_\nu \equiv \int \psi^\dgr_\nu(\br) \psi^\dgr_\nu(\br')
 \Phi(\br - \br') \psi_\nu(\br') \psi_\nu(\br) \; d\br d\br' \; ,
\ee
related to the potential energy part of the Hamiltonian (\ref{82}),
we have for the latter
\be
\label{85}
\hat H_\nu = w_\nu \hat K_\nu +
\frac{w_\nu^2}{2} \; \hat\Phi_\nu \; .
\ee

In this way, we have derived all basic equations for treating
heterophase systems. The derivation has been based on the following
three major points making it possible to separate different thermodynamic
phases. First, to distinguish the phases in the space of microscopic states,
the notion of weighted Hilbert spaces is introduced. Second, to separate the
phases in real space, the manifold indicators were employed. And, finally,
the procedure of averaging over phase configurations is accomplished, leading
to the set of equations for equilibrium on average phase replicas. The idea
of the averaging procedure reminds the method of averaging \cite{80} and the
scale separation approach \cite{81,82,83}, used for nonlinear equations in
dynamical theory. The main difference from the latter is that here we have
averaged out slow heterophase fluctuations, slow with respect to the fast
microscopic motion of particles, while in dynamical theory one usually
averages out fast fluctuations, leaving at the end the slow motion of
guiding centers.

\section{Stability Conditions}

The developed theory should be complimented by an important addition
discussing the stability of heterophase systems. Thermodynamic stability
is characterized by the minimization of a thermodynamic potential. For
concreteness, let us consider the case of a heterophase system, where two
phases coexist, so that $\nu = 1,2$. And let us denote
\be
\label{86}
w_1 \equiv w \; , \qquad w_2 = 1 - w \; .
\ee
Minimizing the thermodynamic potential (\ref{79}) with respect to $w$ implies
\be
\label{87}
\frac{\prt y}{\prt w} = 0 \; , \qquad
\frac{\prt^2 y}{\prt w^2} \; > \; 0 \; .
\ee
The first of these equations gives
\be
\label{88}
\lgl \; \frac{\prt\wdH}{\prt w} \; \rgl \; = \; 0 \; .
\ee
And the second condition leads to the inequality
\be
\label{89}
\lgl \; \frac{\prt^2\wdH}{\prt w^2} \; \rgl \; > \;
\bt \lgl \left ( \frac{\prt\wdH}{\prt w} \right )^2 \rgl \; .
\ee
Since the right-hand side in the above inequality is nonnegative,
the necessary condition of heterophase stability is
\be
\label{90}
\lgl \; \frac{\prt^2\wdH}{\prt w^2} \; \rgl \; > \; 0 \; .
\ee

To specify these conditions, let us take the effective Hamiltonian (\ref{78}),
with the replica Hamiltonians (\ref{85}). And let us use the notations for the
averages
\be
\label{91}
K_\nu \equiv \lgl \hat K_\nu \rgl \; , \qquad
\Phi_\nu \equiv \lgl \hat\Phi_\nu \rgl \; .
\ee
Then we have
$$
\frac{1}{N} \; \lgl \; \frac{\prt\wdH}{\prt w} \; \rgl =
K_1 + w \Phi_1 - K_2 - ( 1 - w ) \Phi_2 \;
$$
and
$$
\frac{1}{N} \; \lgl \; \frac{\prt^2\wdH}{\prt w^2} \; \rgl =
\Phi_1 + \Phi_2 \; .
$$
This yields the equation for the phase probability
\be
\label{92}
w = \frac{\Phi_2 + K_2 - K_1}{ \Phi_1 + \Phi_2} \; .
\ee
The stability condition (\ref{89}) gives
\be
\label{93}
\Phi_1 + \Phi_2 \; > \; \frac{\bt}{N} \;
\lgl \left ( \frac{\prt\wdH}{\prt w} \right )^2 \rgl \; ,
\ee
and from the stability condition (\ref{90}), we get
\be
\label{94}
\Phi_1 + \Phi_2 \; > \; 0 \; .
\ee

One more condition, follows from the definition of $w_\nu$ as of phase
probabilities, according to which
$$
0 \; \leq w \; \leq 1 \; .
$$
This, with the use of (\ref{92}), results in the inequalities
\be
\label{95}
- \Phi_1 \; \leq K_1 - K_2  \; \leq \Phi_2  \; .
\ee

The stability conditions must be valid in order that the considered
heterophase system could really exist. Note that the stability condition
(\ref{90}) is analogous to the condition of diffusion stability \cite{84}.

\section{Heterophase Ferroelectrics}

To illustrate how the developed theory works, let us consider ferroelectric
materials. Ferroelectrics are known to be a good example of matter
demonstrating heterophase properties. In many ferroelectrics, above the
transition point from the disordered into the ordered state, there exist
small polarized clusters \cite{37}. These fluctuational embryos of the
ordered phase inside the disordered phase where termed by Cook \cite{36}
antiphase fluctuations, emphasizing that they where a particular case of
heterophase fluctuations, whose existence was predicted by Frenkel \cite{85}.
Such embryos of one phase inside another phase can exist in a whole region
around the phase-transition point $T_c$. At temperatures below $T_c$, the
embryos of the paraelectric phase inside the ferroelectric phase arise at
a temperature $T_n$ called the {\it lower nucleation point} \cite{86}. And
above $T_c$, there is another temperature $T_n^*$, called the {\it upper
nucleation point} \cite{86}, where the embryos of the ferroelectric phase
appear inside the paraelectric phase. Thus, around the phase transition
temperature, there can exist a region of temperatures
$$
T_n < T_c < T_n^*  \; ,
$$
where the embryos of two phases coexist. The phase transition, generically,
can be either of first or of second order, but the appearance of heterophase
fluctuations, usually smears it into a continuous crossover \cite{41,42,43,44}.

Heterophase fluctuations were observed, for instance, in such well known
ferroelectrics as HCl, DCl, mixed crystals HCl$_{1-x}$DCl$_x$, and RbCaF$_3$,
where they were intensively studied by Brookeman and Rigamonti using nuclear
quadrupole resonance \cite{42,43} and nuclear magnetic resonance \cite{44}.
Heterophase fluctuations in these ferroelectrics arise in a finite region around
the phase transition point. The appearance of these fluctuations occurs even
without external defects, although the presence of defects intensifies their
nucleation \cite{41}.

Such heterophase fluctuations have also been observed in many other
ferroelectrics, e.g., in C$_4$O$_4$H$_2$ \cite{41}, in KH$_2$As$_4$ \cite{46},
in Rb$_x$(ND$_4$)$_{1-x}$D$_2$PO$_4$ \cite{47}, in Na$_x$Bi$_{1-x}$TiO$_3$
\cite{49}, and others. They also arise in such novel materials as relaxor
ferroelectrics, for example, in (PbZn$_x$Nb$_{1-x}$O$_3$) \cite{55}, in
Ba$_2$NdTi$_2$Nb$_3$O$_{15}$ and Ba$_2$La$_{0.5}$Nd$_{0.5}$Ti$_2$Nb$_3$O$_{15}$
\cite{56,58}, and, with a high probability, in many other relaxors, such as
PbMg$_{1/3}$Nb$_{2/3}$O$_3$,
(Ba$_x$Pb$_{1-x}$)(Zn$_{1/3}$Nb$_{2/3}$)O$_3$,
(Sr$_x$Pb$_{1-x}$)(Zn$_{1/3}$Nb$_{2/3}$)O$_3$,
(Ba$_x$Pb$_{1-x}$)(Yb$_{0.5}$Nb$_{0.5}$)O$_3$,
Pb$_{1-x}$Ba$_x$(Yb$_{1/2}$Ta$_{1/2}$)O$_3$,
(Fe$_{1/2}$Nb$_{1/2}$)O$_3$, Pb(Fe$_{1/2}$Ta$_{1/2}$)O$_3$,
Pb(Yb$_{1/2}$Ta$_{1/2}$)O$_3$-Pb(Fe$_{1/2}$Ta$_{1/2}$)O$_3$,
and Pb(Mg$_{1/3}$Nb$_{2/3}$)O$_3$-PbTiO$_3$, \cite{58}.

A simple model of a heterophase ferroelectric has been considered in
Refs. \cite{50,87,88}. In this model, phonon degrees of freedom were
not taken into account. However, the latter are important because of
two reasons. First, the appearance of heterophase fluctuations is
frequently accompanied by heterostructural fluctuations \cite{89,90},
which are connected with phonon excitations. Second, the occurrence of
the paraelectric-ferroelectric phase transition is intimately related
with phonon characteristics that can be directly measured. In the
present paper, we derive a generalized model of a heterophase
ferroelectric, taking into account the phonon degrees of freedom.
This makes it possible to find out the influence of the heterophase
fluctuations on such observable quantities as the Debye-Waller factor
and sound velocity. This influence is especially pronounced in the
phase-transition region.

\section{Basic Hamiltonian}

Let us consider a ferroelectric, in which there can arise the embryos
of the competing phase. So that the sample can house two coexisting
phases, ferroelectric and paraelectric, being randomly intermixed with
each other. The ferroelectric phase will be indexed by $\nu=1$ and
the paraelectric phase will be labeled by the index $\nu=2$.  For
concreteness, we shall study the model of a ferroelectric, in which
the order is characterized by the pseudospin operator $S_j^z$ describing
the shift of a charged particle into one of the wells of a double-well
potential at the cite $j$ of the crystalline lattice \cite{91,92}. Then
the ordered ferroelectric phase corresponds
to the nonzero order parameter
\be
\label{96}
\lgl S_{j1}^z \rgl \neq 0 \; .
\ee
On the contrary, the disordered paraelectric phase is characterized by
the zero order parameter
\be
\label{97}
\lgl S_{j2}^z \rgl \equiv 0 \; .
\ee

Starting with the ferroelectric Hamiltonian, having the mathematical structure
characterized by the pseudospin variables \cite{91,92}, we follow the general
scheme described above, and after averaging over the random phase configurations,
we come to the effective Hamiltonian (\ref{78}) with the replica Hamiltonians
$$
\hat H_\nu = w_\nu \sum_j \; \frac{\bp_j^2}{2m} \; + \;
\frac{w_\nu^2}{2} \; \sum_{i\neq j} A(\br_{ij}) \; - \;
w_\nu \Om \sum_j S_{j\nu}^x \; +
$$
\be
\label{98}
+ \; w_\nu^2 \sum_{i\neq j} B(\br_{ij}) S_{i\nu}^x S_{j\nu}^x \; - \;
w_\nu^2 \sum_{i\neq j} I(\br_{ij}) S_{i\nu}^z S_{j\nu}^z \; .
\ee
Here, the first term represents kinetic energy, $A(\bf r), B(\bf r)$, and $I(\bf r)$
are particle interactions, $\Omega$ is the tunneling frequency, and the abbreviated
notation
\be
\label{99}
\br_{ij} \equiv \br_i - \br_j \; .
\ee
is employed.

The phonon variables can be introduced in the standard way by defining the deviation
$\bf u_j$ from the lattice site with the lattice vector $\bf a_j$ as
\be
\label{100}
\br_j = \ba_j + \bu_j \; .
\ee
The lattice vectors are assumed to form an equilibrium lattice, being defined as
the averages
\be
\label{101}
\ba_j \equiv \lgl \br_j \rgl \; .
\ee
Hence the average deviation, by definition, is zero,
\be
\label{102}
\lgl \bu_j \rgl = 0 \; .
\ee

The interactions, as usual, are supposed to be symmetric with respect to the spatial
coordinate inversion:
\be
\label{103}
A(-\br_{ij}) = A(\br_{ij}) \; , \qquad
B(-\br_{ij}) = B(\br_{ij}) \; , \qquad
I(-\br_{ij}) = I(\br_{ij}) \; .
\ee
In what follows, we shall also use the short-hand notation for the vector differences
\be
\label{104}
\ba_{ij} \equiv \ba_i - \ba_j \; .
\ee
and
\be
\label{105}
\bu_{ij} \equiv \bu_i - \bu_j \; .
\ee

The interactions are expanded in powers of the deviations as
\be
\label{106}
A(\br_{ij}) \cong A_{ij} + \sum_\al
A_{ij}^\al u_{ij}^\al \; + \; \frac{1}{2} \; \sum_{\al\bt}
A_{ij}^{\al\bt} u_{ij}^\al u_{ij}^\bt \; ,
\ee
where
\be
\label{107}
A_{ij} \equiv A(\ba_{ij}) \; , \qquad
A_{ij}^\al \equiv \frac{\prt A_{ij}}{\prt a_i^\al} \; ,
\qquad A_{ij}^{\al\bt} \equiv
\frac{\prt^2 A_{ij}}{\prt a_i^\al \prt a_i^\bt} \; .
\ee
The same expansions are made for $B(\bf r)$ and $I(\bf r)$. Owing to the
symmetry properties (\ref{103}), we have
\be
\label{108}
A_{ij} \equiv A_{ji} \; , \qquad A_{ij}^\al = - A_{ji}^\al \; ,
\qquad
A_{ij}^{\al\bt} = A_{ij}^{\bt\al} = A_{ji}^{\bt\al} = A_{ji}^{\al\bt} \; .
\ee

In the Hamiltonian (\ref{98}), the double summation over the lattice excludes
the self-action terms with $i=j$. In order to simplify the notation, we can sum
over all lattice sites, setting the diagonal elements
\be
\label{109}
A_{ii} = B_{ii} = I_{ii}  \equiv 0 \; .
\ee
The lattice is treated as ideal, because of which
\be
\label{110}
A \equiv \sum_j A_{ij} = const \;
\ee
does not depend on the index $i$. Using the ideality of the lattice, we get
\be
\label{111}
\sum_j A_{ij}^\al = \frac{\prt A}{\prt a_i^\al} = 0 \; .
\ee
and
\be
\label{112}
\sum_j A_{ij}^{\al\bt} =
\frac{\prt^2 A}{\prt a_i^\al \prt a_i^\bt} = 0 \; .
\ee
Consequently,
\be
\label{113}
\sum_{ij} A_{ij}^\al u_{ij}^\al = 0 \; .
\ee

The same type of expansions is accomplished for both the ferroelectric and
paraelectric phases. Therefore, in what follows, we shall consider the resulting
transformations only for the ferroelectric phase, keeping in mind that the same is
done for the paraelectric phase. And to simplify the notation, we shall not write
explicitly the index $\nu=1$. Then, substituting the above expansions into
Hamiltonian (\ref{98}), with $\nu=1$, and invoking the notation
\be
\label{114}
S_{ij}^\al \equiv S_i^\al S_j^\al \; ,
\ee
we obtain the Hamiltonian
$$
\hat H_1 = w \sum_j \; \frac{\bp^2_j}{2m} \; + \; \frac{w^2}{2} N A  \; + \;
\frac{w^2}{4} \sum_{ij}
\sum_{\al\bt} A_{ij}^{\al\bt} u_{ij}^\al u_{ij}^\bt \; - \;
w \Om \sum_j S_j^x \; +
$$
$$
+ \; w^2 \sum_{ij} \left ( B_{ij} + \sum_\al B_{ij}^\al u_{ij}^\al +
\frac{1}{2} \sum_{\al\bt} B_{ij}^{\al\bt} u_{ij}^\al u_{ij}^\bt
\right ) S_{ij}^x \; -
$$
\be
\label{115}
- \; w^2 \sum_{ij} \left ( I_{ij} + \sum_\al I_{ij}^\al u_{ij}^\al +
\frac{1}{2} \sum_{\al\bt} I_{ij}^{\al\bt} u_{ij}^\al u_{ij}^\bt
\right ) S_{ij}^z \; ,
\ee
in which $w \equiv w_1$.

\section{Pseudospin-Phonon Decoupling}

The obtained Hamiltonian is yet too complicated to be treated, and some approximation
is required. For any approximation, we have to keep in mind that
\be
\label{116}
\lgl \; u_{ij}^\al \; \rgl = \lgl \; u_i^\al \; \rgl \; - \;
\lgl \; u_j^\al \; \rgl \; = \; 0 \; .
\ee

Because the phonon and pseudospin operators are of different nature, it is reasonable
to decouple them in the second-order, with respect to the deviations, terms as
\be
\label{117}
u_{ij}^\al u_{ij}^\bt S_{ij}^\gm  \; =  \;
\lgl \; u_{ij}^\al u_{ij}^\bt \; \rgl \; S_{ij}^\gm \; + \;
u_{ij}^\al u_{ij}^\bt \; \lgl \; S_{ij}^\gm \; \rgl \; - \;
\lgl \; u_{ij}^\al  u_{ij}^\bt \; \rgl \;
\lgl \; S_{ij}^\gm \; \rgl \; .
\ee
At the same time, the terms linear in the deviations can be left for a while, since
later they can be dealt with by using canonical transformations.

Using again the lattice ideality, we see that
\be
\label{118}
\sum_{ij} A_{ij}^{\al\bt} u_{ij}^\al u_{ij}^\bt =
2 \sum_{ij} A_{ij}^{\al\bt} u_i^\al u_j^\bt \; .
\ee
The average $<S_{ij}>$ depends on the difference $\bf a_{ij}$. Thence
$$
\sum_j B_{ij}^{\al\bt} \; \lgl \; S_{ij}^x \; \rgl \; = \;
\frac{\prt^2}{\prt a_i^\al \prt a_i^\bt}\;
\sum_j B_{ij} \; \lgl \; S_{ij}^x \; \rgl \; = \; 0 \; ,
$$
\be
\label{119}
\sum_j I_{ij}^{\al\bt} \; \lgl \; S_{ij}^z \; \rgl \; = \;
\frac{\prt^2}{\prt a_i^\al \prt a_i^\bt}\;
\sum_j I_{ij} \; \lgl \; S_{ij}^z \; \rgl \; = \; 0 \; .
\ee
Similarly,
$$
\sum_{ij} B_{ij}^{\al\bt} u_{ij}^\al u_{ij}^\bt \;
\lgl \; S_{ij}^x \; \rgl \;  = \; 2 \sum_{ij} B_{ij}^{\al\bt} \;
\lgl \; S_{ij}^x \; \rgl \; u_i^\al u_j^\bt \; ,
$$
\be
\label{120}
\sum_{ij} I_{ij}^{\al\bt} u_{ij}^\al u_{ij}^\bt \;
\lgl \; S_{ij}^z \; \rgl \;  = \; 2 \sum_{ij} I_{ij}^{\al\bt} \;
\lgl \; S_{ij}^z \; \rgl \; u_i^\al u_j^\bt \; .
\ee

Let us introduce the notation for the {\it dynamical matrix}
\be
\label{121}
\Phi_{ij}^{\al\bt} \equiv  A_{ij}^{\al\bt} +
2 B_{ij}^{\al\bt} \; \lgl \; S_{ij}^x \; \rgl \; - \;
2 I_{ij}^{\al\bt} \; \lgl \; S_{ij}^z \; \rgl \;
\ee
and for the renormalized interactions
\be
\label{122}
\widetilde B_{ij} \equiv B_{ij} \; + \; \sum_{\al\bt} B_{ij}^{\al\bt} \;
\lgl \; u_i^\al u_j^\bt \; \rgl \; .
\ee
and
\be
\label{123}
\widetilde I_{ij} \equiv I_{ij} \; + \; \sum_{\al\bt} I_{ij}^{\al\bt} \;
\lgl \; u_i^\al u_j^\bt \; \rgl \; .
\ee

Separating the non-operator energy part
\be
\label{124}
E_1 \equiv \frac{w^2}{2} \; N A \; + \; w^2 \sum_{ij} \sum_{\al\bt}
\left ( I_{ij}^{\al\bt} \; \lgl \; S_{ij}^z \; \rgl \; - \;
B_{ij}^{\al\bt} \; \lgl S_{ij}^x \; \rgl \right )
\lgl \; u_i^\al u_j^\bt \; \rgl \; ,
\ee
we reduce Hamiltonian (\ref{115}) to the sum
\be
\label{125}
\hat H_1 = E_1 + \hat H_{ph} + \hat H_{ps} + \hat H_{lin} \; .
\ee
The second term here is the effective phonon Hamiltonian
\be
\label{126}
\hat H_{ph} = w \sum_j \; \frac{\bp^2_j}{2m} \; + \;
\frac{w^2}{2} \; \sum_{ij}
\sum_{\al\bt} \Phi_{ij}^{\al\bt} u_i^\al u_j^\bt \; .
\ee
The third term is the effective pseudospin Hamiltonian
\be
\label{127}
\hat H_{ps} = - w \Om \sum_j S_j^x \; + \; w^2 \sum_{ij}
\left ( \widetilde B_{ij} S_{ij}^x \; - \; \widetilde I_{ij} S_{ij}^z
\right ) \; .
\ee
And the last term is the linear pseudospin-phonon interaction Hamiltonian.
The latter is obtained by invoking the properties
$$
\sum_{ij} B_{ij}^\al u_{ij}^\al S_{ij}^x = 2 \sum_{ij} B_{ij}^\al
S_{ij}^x u_i^\al = - 2 \sum_{ij} B_{ij}^\al S_{ij}^x u_j^\al \; ,
$$
\be
\label{128}
\sum_{ij} I_{ij}^\al u_{ij}^\al S_{ij}^z = 2 \sum_{ij} I_{ij}^\al
S_{ij}^z u_i^\al = - 2 \sum_{ij} I_{ij}^\al S_{ij}^z u_j^\al \;
\ee
and denoting
\be
\label{129}
K_{ij}^\al \equiv B_{ij}^\al S_{ij}^x \; - \;
I_{ij}^\al S_{ij}^z \; ,
\ee
which yields
\be
\label{130}
\hat H_{lin} = -
2 w^2 \sum_{ij} \sum_\al K_{ij}^\al u_j^\al \; .
\ee
Thus, the Hamiltonian parameters are renormalized due to the interactions between
pseudospins and phonons.

\section{Dressed Phonons}

The quantization of the phonon degrees of freedom can be done in the way, similar to
how it is done for pure crystalline phases. However, here we have to be careful,
taking into account the presence of the factors $w=w_1$, characterizing  the weight
of the related phase. Thus, the eigenproblem  equation for the phonon frequencies and
polarization vectors takes the form
\be
\label{131}
\frac{w}{m} \; \sum_{\al\bt} \Phi_{ij}^{\al\bt}
\exp\left ( i\bk \cdot \ba_{ij} \right ) e^\bt_{ks} =
\om_{ks}^2 e^\al_{ks} \; .
\ee
The phonon frequencies and polarization vectors can be chosen to be symmetric with
respect to the momentum inversion,
$$
\om_{-ks} = \om_{ks} \; , \qquad \bfe_{-ks} = \bfe_{ks} \; .
$$
The polarization vectors are orthonormalized, such that
$$
\bfe_{ks} \cdot \bfe_{ks'} = \dlt_{ss'} \; , \qquad
\sum_s e_{ks}^\al e_{ks}^\bt = \dlt_{\al\bt} \; ,
$$
with the momentum summation over the Brillouin zone. The eigenproblem (\ref{131})
can be rewritten as
\be
\label{132}
\frac{w}{m} \; \sum_\bt \Phi_k^{\al\bt} e_{ks}^\bt =
\om_{ks}^2 e_{ks}^\al \; ,
\ee
where
$$
\Phi_k^{\al\bt} \equiv \sum_j \Phi_{ij}^{\al\bt}
e^{ i\bk \cdot \ba_{ij} } \; .
$$

The phonon quantization, in the presence of the linear pseudospin-phonon interactions,
differs from the standard case by the necessity to involve a nonuniform canonical
transformation
$$
\bp_j = -\; \frac{i}{\sqrt{N}} \;
\sum_{ks} \; \sqrt{ \frac{m}{2}\; \om_{ks} } \;\;
\bfe_{ks} \left ( b_{ks} - b_{-ks}^\dgr \right )
e^{i\bk \cdot\ba_j} \; ,
$$
\be
\label{133}
\bu_j = {\bf v}_j + \frac{1}{\sqrt{N}} \;
\sum_{ks} \; \frac{\bfe_{ks}}{
\sqrt{ 2m \om_{ks} } } \;
\left ( b_{ks} + b_{-ks}^\dgr \right )
e^{i\bk \cdot\ba_j} \; .
\ee
The nonuniformity comes through an additional term in the expression for $u_j$.

Hamiltonian (\ref{125}) transforms to
\be
\label{134}
\hat H_1 = E_1 + \hat H_{ph}' + \hat H_{ps} +
\hat H_{lin}' \; .
\ee
The effective phonon Hamiltonian is
\be
\label{135}
\hat H_{ph}' = w \sum_{ks} \om_{ks}
\left ( b_{ks}^\dgr b_{ks} + \frac{1}{2} \right ) \; ,
\ee
with the phonon frequency given by the equation
\be
\label{136}
\om_{ks}^2 = \frac{w}{m} \; \sum_j \sum_{\al\bt} \;
\Phi_{ij}^{\al\bt} e_{ks}^\al e_{ks}^\bt \;
e^{ i\bk \cdot \ba_{ij} } \; .
\ee
The latter equation can also be written as
$$
\om_{ks}^2 = \frac{w}{m} \; \sum_{\al\bt} \;
\Phi_k^{\al\bt} e_{ks}^\al e_{ks}^\bt \; .
$$
The momentum $\bf k$ pertains to the Brillouin zone.

After transformation (\ref{133}), the term, remaining from the renormalized
linear pseudospin-phonon interaction, reads as
\be
\label{137}
\hat H_{lin}' = -2 w^2 \sum_{ij} \sum_\al \;
K_{ij}^\al v_j^\al \; ,
\ee
with
\be
\label{138}
v_f^\al = \frac{1}{2N} \; \sum_{ij} \sum_\bt \;
\gm_{jf}^{\al\bt} K_{ij}^\bt \; ,
\ee
where the notation
\be
\label{139}
\gm_{ij}^{\al\bt} \equiv 4w \sum_{ks} \;
\frac{e_{ks}^\al e_{ks}^\bt }{m\om_{ks}^2} \;
\exp \left ( i\bk \cdot \ba_{ij} \right ) \;
\ee
is used. The latter quantity possesses the properties
$$
\gm_{ij}^{\al\bt} = \gm_{ji}^{\al\bt} =
\gm_{ji}^{\bt\al} = \gm_{ij}^{\bt\al} \; .
$$
Combining (\ref{137}) and (\ref{138}) gives
\be
\label{140}
\hat H_{lin}' = - \; \frac{w^2}{N} \;
\sum_{ij} \sum_{fg} \sum_{\al\bt} \;
K_{ij}^\al \gm_{jf}^{\al\bt} K_{fg}^\bt \; ,
\ee
which shows that this is an effective four-pseudospin interaction.

In the summation over momenta in (\ref{139}), the main contribution comes
from the term with $k=0$ because of the fast oscillations of the exponential.
Therefore, expression (\ref{139}) can be well approximated as
\be
\label{141}
\gm_{ij}^{\al\bt} \cong \gm^{\al\bt} \equiv 4w \sum_{ks} \;
\frac{e_{ks}^\al e_{ks}^\bt}{m\om_{ks}^2 } \; .
\ee
Then (\ref{138}) becomes
\be
\label{142}
v_f^\al \; \cong \; \frac{1}{2N} \; \sum_{ij} \sum_\bt \;
\gm^{\al\bt} K_{ij}^\bt \; .
\ee
Owing to the property
$$
\sum_{ij} \; K_{ij}^\bt \; = \;
- \sum_{ij} \; K_{ij}^\bt = 0 \; ,
$$
we have
\be
\label{143}
\hat H_{lin}' = 0 \; , \qquad v_f^\al = 0 \; .
\ee
Consequently, the term $H_{lin}'$ in Hamiltonian (\ref{134}) can be omitted.

This is in agreement with the following. The existence of the linear,
in the deviations, terms in Hamiltonian (\ref{130}), formally, leads to the fact that,
according to (\ref{133}), the average $<u_j>$ could be nonzero, which, however would be
in contradiction with condition (\ref{102}). Therefore, the linear in deviations terms
should be zero, as in Eqs. (\ref{143}). Also, the appearance of the linear terms in
Hamiltonian (\ref{130}) breaks the symmetry of the initial Hamiltonian with respect
to the inversion $u_j \rightarrow -u_j$. Generally, such linear terms should either
be zero, due to symmetry properties, or are to be canceled by counterterms preserving
the equilibrium condition (\ref{102}).

The same conclusion could be obtained, if, when decoupling the pseudospin and
phonon degrees of freedom, we would employ the decoupling
\be
\label{144}
u_{ij}^\al S_{ij}^\bt \; = \;
\lgl \; u_{ij}^\al \; \rgl S_{ij}^\bt \; + \;
u_{ij}^\al \lgl \; S_{ij}^\bt \; \rgl \; - \;
\lgl \; u_{ij}^\al \; \rgl
\lgl \; S_{ij}^\bt \; \rgl \; ,
\ee
which, in view of (\ref{116}), yields
\be
\label{145}
u_{ij}^\al S_{ij}^\bt \; = \;
u_{ij}^\al \; \lgl \; S_{ij}^\bt \; \rgl \; .
\ee
Then, employing the properties
$$
\sum_j \; B_{ij}^\al \; \lgl \; S_{ij}^x \; \rgl \; = \;
\frac{\prt}{\prt a_i^\al} \; \sum_j \;
B_{ij} \; \lgl \; S_{ij}^x \; \rgl \; = \; 0 \; ,
$$
\be
\label{146}
\sum_j \; I_{ij}^\al \; \lgl \; S_{ij}^z \; \rgl \; = \;
\frac{\prt}{\prt a_i^\al} \; \sum_j \;
I_{ij} \; \lgl \; S_{ij}^z \; \rgl \; = \; 0 \;
\ee
and
\be
\label{147}
\sum_{ij} B_{ij}^\al u_{ij}^\al \;
\lgl \; S_{ij}^x \; \rgl \; = \; 0 \; , \qquad
\sum_{ij} I_{ij}^\al u_{ij}^\al \;
\lgl \; S_{ij}^z \; \rgl \; = \; 0 \; ,
\ee
following from the ideality of the lattice, results in the Hamiltonian
\be
\label{148}
\hat H_1 = E_1 + \hat H_{ph} + \hat H_{ps} \; .
\ee
The first term here is the nonoperator part (\ref{124}), the second term is
the effective phonon Hamiltonian
\be
\label{149}
\hat H_{ph} = w \sum_{ks} \om_{ks} \left ( b_{ks}^\dgr b_{ks} +
\frac{1}{2} \right ) \; .
\ee
And the effective pseudospin Hamiltonian is
\be
\label{150}
\hat H_{ps} = - w \Om \sum_j S_j^x \; + \;
w^2 \sum_{ij} \widetilde B_{ij} S_i^x S_j^x \; - \;
w^2 \sum_{ij} \widetilde I_{ij} S_i^z S_j^z \; .
\ee

The pseudospin interactions are renormalized by the existence of the phonon
vibrations. And the phonon characteristics are renormalized due to the phonon
interactions with pseudospins. In addition, all quantities are renormalized by
the presence of the heterophase fluctuations. It is, therefore, possible to call
the resulting effective phonons as {\it dressed phonons}.

\section{Heterophase Fluctuations}

With the derived effective Hamiltonian (\ref{148}), we can explicitly calculate
all phonon characteristics. For instance, the phonon distribution is
\be
\label{151}
n_{ks} \; \equiv \;
\lgl \; b_{ks}^\dgr b_{ks} \; \rgl \; = \;
\left [ \exp \left ( \frac{w\om_{ks}}{T} \right ) -1
\right ]^{-1} \; .
\ee
The deviation-deviation correlation function read as
\be
\label{152}
\lgl \; u_j^\al u_j^\bt \; \rgl \; = \; \frac{1}{2N} \;
\sum_{ks} \; \frac{e_{ks}^\al e_{ks}^\bt}{m\om_{ks} } \;
\coth \left (\frac{w\om_{ks}}{2T} \right ) \; .
\ee
The mean kinetic energy becomes
\be
\label{153}
\lgl \; \frac{\bp^2}{2m} \; \rgl \; = \; \; \frac{1}{4N} \;
\sum_{ks} \om_{ks} \coth \left (\frac{w\om_{ks}}{2T} \right ) \; .
\ee

In order to treat the pseudospin variables, we can use the mean-field
approximation for the products $S_i^\alpha S_j^\beta$. This is possible because
the critical region in ferroelectrics is known \cite{91,92} to be narrow due
to the smallness of the Ginzburg number. Then, in the mean-field approximation
for the pseudospins, we find the averages for the $x$-component
\be
\label{154}
\lgl \; S_j^x \; \rgl \; = \; w \; \frac{\Om_j}{2H_j} \;
\tanh \left (\frac{H_j}{2T} \right ) \; ,
\ee
for the $y$-component
\be
\label{155}
\lgl \; S_j^y \; \rgl \; = \; 0\; ,
\ee
and for the $z$-component
\be
\label{156}
\lgl \; S_j^z \; \rgl \; = \; w^2 \lgl \; S_j^z \; \rgl \;
\frac{\widetilde I}{H_j} \;
\tanh \left (\frac{H_j}{2T} \right ) \; .
\ee
Here, the notations are used for the effective tunneling frequency
\be
\label{157}
\Om_j = \Om - 2w \widetilde B \;
\lgl \; S_j^x \; \rgl \;
\ee
and the effective field
\be
\label{158}
H_j = w \;
\sqrt{\Om_j^2 + 4w^2 \widetilde I^2 \; \lgl \; S_j^z \; \rgl}~,
\ee
in which
\be
\label{159}
\widetilde B \equiv \sum_j \; \widetilde B_{ij} \; , \qquad
\widetilde I \equiv \sum_j \; \widetilde I_{ij} \; .
\ee

To deal further with considering the properties of the heterophase ferroelectric,
we need to restore the phase indices $\nu = 1,2$. The effective Hamiltonian of
the heterophase system is
\be
\label{160}
\widetilde H = \hat H_1 \; \bigoplus \; \hat H_2 \; .
\ee
The conditions, distinguishing the ordered and disordered phases are
$$
\lgl \; S_{j1}^z \; \rgl \; \neq \; 0 \; , \qquad
\lgl \; S_{j2}^z \; \rgl \; = \; 0 \; .
$$
All necessary equations for the ferroelectric phase are written above.
The equations for the paraelectric phase can be obtained from the above
expressions by setting there the order parameter $<S_j^z> = 0$. The equations
for the phase probabilities are given in Sec. 8.

\section{Phase Transition}

The total system of equations, defining the heterophase ferroelectric, can be solved
numerically. Below, we present some of the most interesting conclusions describing the
influence of the heterophase fluctuations on the system properties. The strongest
influence of these fluctuations occurs in the vicinity of the phase transition point,
that is represented by the temperature
\be
\label{161}
\widetilde T_c \; = \;
\frac{(1-\widetilde b)\widetilde\om}{4{\rm artanh}(2\widetilde\om) } \; ,
\ee
where
$$
\widetilde b \; \equiv \;
\frac{\widetilde B}{\widetilde I + \widetilde B} \; ,
\qquad
\widetilde\om \; \equiv \;
\frac{\widetilde\Om}{\widetilde I + \widetilde B} \; .
$$
The existence of the heterophase fluctuations can be noticed and their influence
measured by studying, e.g., the Debye-Waller factor $f_{DW}$ and sound velocity $s$.
It is convenient to characterize the influence of the heterophase fluctuations by
comparing their values in the presence of the latter with the related values without
these fluctuations. For example, comparing the sound velocity $s$ in the heterophase
system with the sound velocity $s_0$ in a pure system without such fluctuations, it
is useful to introduce the relative sound-velocity decrease
\be
\label{162}
\dlt s \; \equiv \; \frac{s-s_0}{s_0} \; .
\ee

Another useful characteristic is the Debye-Waller factor $f_{DW}$ that can be measured
by x-ray scattering, coherent neutron scattering, and by M\"{o}ssbauer spectroscopy. All
details and definitions can be found in the books \cite{93,94,95}. We need to compare
the Debye-Waller factor $\widetilde f_{DW}$ for the heterophase system and its value
$f_{DW}$ for a pure system. Again, it is useful to employ the relative Debye-Waller
factor decrease
\be
\label{163}
\dlt f_{DW} \; \equiv \;
\frac{\widetilde f_{DW} - f_{DW} }{f_{DW} } \; .
\ee

It is interesting that, as follows from numerical calculations, these quantities, the
relative sound velocity decrease and the relative Debye-Waller factor decrease, are
universal, weakly depending on the considered materials. For the relative sound velocity
decrease at the critical transition point, we get
\be
\label{164}
\dlt s \approx -0.3 \qquad ( T = T_c) \;
\ee
and for the relative Debye-Waller factor decrease at this point,
\be
\label{165}
\dlt f_{DW} \approx -0.3 \; .
\ee
This decrease of the sound velocity and of the Debye-Waller factor is
due to the scattering caused by heterophase fluctuations.

\section{Conclusion}

The systems are addressed, exhibiting phase transitions between thermodynamic phases
with different symmetry, in which spontaneous symmetry breaking can be accompanied
by local spontaneous symmetry restoration, caused by the appearance of heterophase
fluctuations. In such systems, above the phase transition point, where a disordered
phase dominates, there appear the germs of the ordered phase. And below the transition
point, where an ordered phase prevails, there arise the embryos of the disordered phase.
Thus, around the phase transition point there is a region, where the phases with higher
and lower symmetries coexist. The embryonic regions of a competing phase are distributed
randomly in space. Their characteristic sizes are mesoscopic, so that the typical size
$r_f$  of such an embryo is much larger than the mean interparticle distance $a$, but
much smaller than the characteristic size of the whole system $L$, that is,
$$
  a \ll r_f \ll L  \; .
$$
The typical size $r_f$ should be understood as an average size, since the shapes of
the embryonic heterophase fluctuations are not necessarily regular, but can be rather
ramified. Therefore, terming these germs mesoscopic could be better applied to the average
number of particles $N_f$ composing each of such embryos, as compared to the total
number $N$ of particles in the system, so that
$$
 1 \ll N_f \ll N  \;  .
$$
This type of heterophase fluctuations is common for a number of substances.

The general approach for treating these heterophase systems is developed, being based
on the notion of weighted Hilbert spaces. The real-space distribution of the phases
is described by means of manifold indicators. The averaging over random phase
configurations reduces the problem to the consideration of an effective renormalized
Hamiltonian composed of the phase-replica Hamiltonians representing the phases of
different symmetry. Stability conditions define the geometric phase probabilities
in a self-consistent way.

The method is illustrated by applying it to heterophase ferroelectrics that are the
typical materials exhibiting the appearance of such heterophase fluctuations around
their phase transition points between the paraelectric and ferroelectric phases. The
influence of the heterophase fluctuations is the strongest in the vicinity of the phase
transition point. Numerical calculations show that the occurrence of such fluctuations
leads to the noticeable decrease of the sound velocity and Debye-Walle factor at the
transition point. The {\it relative} values of this decrease turn out to be universal,
only weakly depending on the material parameters.

In conclusion, it is worth mentioning that thermodynamic phases and phase transitions
between them can be conveniently characterized not only by order parameters but also
by order indices \cite{96}. Another important characteristic is the measure of
entanglement in the considered physical system \cite{97}. These three characteristics
are interrelated with each other \cite{98,99,100}. The usual situation is when the
increasing order is accompanied by the decreasing entanglement \cite{100}. Since a
mesoscopic mixture is a system that is between an absolutely disordered and a
completely ordered phases, its entanglement should be between these two limiting
cases. An additional entanglement arises in mesoscopic mixtures because a heterophase
system consists of several spatially separated regions with different symmetry, and
these mesoscopic regions are mutually entangled. This increases the system entanglement,
as compared to the completely ordered case. The problem of studying the entanglement
level of mesoscopic mixtures is extremely interesting. Since the mixture properties
can be regulated by means of thermodynamic parameters, this gives an additional
possibility of governing the level of entanglement in such mesoscopic mixtures, which
does not exist in the case of pure systems. This novel possibility of regulating
entanglement is really exciting, but it is a separate problem that goes out of the
scope of the present paper. It is clear that the mesoscopic mixtures provide a
nontrivial possibility for regulating the system entanglement, which can be used for
quantum information processing.

\section*{Acknowledgements}

I am grateful to E.P. Yukalova for useful advice. Financial support from the Russian
Foundation for Basic Research is appreciated.

\end{document}